\documentclass[preprint,12pt]{elsarticle}

\usepackage{amssymb}
\usepackage{physics}
\usepackage{graphicx}
\usepackage{xcolor}
\usepackage{float}
\usepackage[utf8]{inputenc}
\usepackage{amsmath}
\usepackage{mathtools}
\usepackage{color,soul}

\begin{document}

\begin{frontmatter}



\title{Calculating the Raman Signal  Beyond Perturbation Theory for a Diatomic Molecular Crystal}


\author[inst1]{Peter I. C. Cooke}

\affiliation[inst1]{organization={School of Physics and Astronomy},
            addressline={The University of Edinburgh}, 
            postcode={EH9 3FD},
            country={UK}}

\author[inst2]{Ioan B. Magdău}
\author[inst1]{Graeme J. Ackland}

\affiliation[inst2]{organization={Engineering Laboratory},
            addressline={University of Cambridge}, 
            postcode={CB2 1PZ},
            country={UK}}

\begin{abstract}
We calculate the eigenstates of a diatomic molecule in a range of model mean-field potentials, and evaluate the evolution of their associated Raman spectra with field strength. We demonstrate that dramatic changes in the appearance of the Raman spectrum for a diatomic molecule occur without any associated change in the symmetry of the surrounding potential. The  limiting cases of the quantum eigenstates correspond, in the classical sense, to free rotation, and libration of well-oriented molecules.  However, there are also many mixed modes which are neither rotons nor librons. The consequence for Raman spectroscopy is a series of complications - the non-harmonic potential splits the Raman active modes, and breaks the selection rules on forbidden modes. The mass-dependence of the various states is different - rotors, oscillators and reorientations have $1/m$, $1/\sqrt{m}$ and weaker mass dependence respectively.  This  may allow one to identify the character of the mode with isotope spectroscopy. However it is complicated by mixed modes and  transitions between two different eigenstates with different character.  We conclude that significant changes in the Raman spectrum of molecular systems are insufficient to demonstrate a phase transition since such changes can also occur in a fixed symmetry potential upon increasing field strength.
\end{abstract}

\begin{graphicalabstract}
\includegraphics[width=\textwidth]{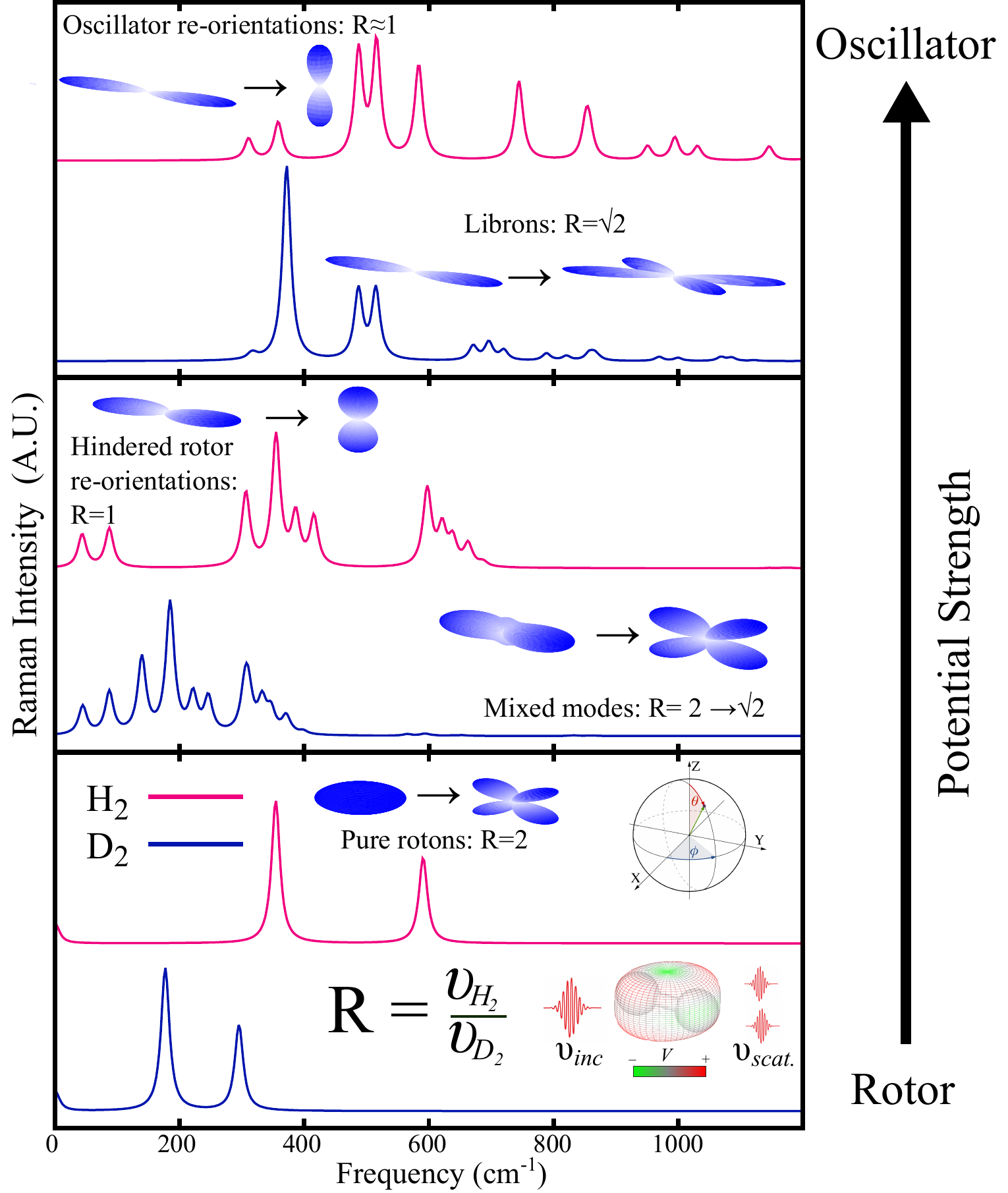}
\end{graphicalabstract}

\begin{highlights}
\item Striking changes in Raman spectra occur  without associated phase transitions.
\item Spectroscopic signature of diatomic molecule is simple in rotor and oscillator limits, but very complex between.
\item Selection rules are valid only in high and low potential limits: for an intermediate strength potential many more modes are spectroscopically allowed.
\end{highlights}


\end{frontmatter}


\section{Introduction}
\label{sec:Introduction}
Optical spectroscopy is one of the most important methods with which one can study solid materials.  The two main methods are Infrared (IR),  in which light is absorbed by the sample, and Raman, in which light is reemitted at a different frequency.  The  energy of the absorbed IR photon, or the shift in energy of the Raman signal, correspond to quantised excitations in the material.

In a free molecule, these quantized excitations are vibrations or rotations.   Selection rules can be derived using group theory.  Each molecular mode contributes equally, so the signal is proportional to the number of molecules, $N$.
In a harmonic, crystalline solid, the excitations are the phonons, and we can apply Bloch's theorem to calculate the phonon spectrum, assigning each normal mode to a reciprocal space $q$-vector.  Conservation of momentum means that the only {\it allowed} modes - those which produce a spectroscopic signal - have $q=0$.  The number of modes is less than three times the number of atoms in the unit cell, but because of constructive interference throughout the crystal each mode is amplified by the total number of molecules, so again these modes produce a signal proportional to the number of molecules, $N$.

Molecular crystals lie between these two extremes.  If the molecules are freely rotating, Bloch's theorem applies only to the molecular centre of mass. So there will be a combination of lattice phonons and free-molecule vibrations and rotations.  If the intermolecular forces are strong enough to inhibit the rotation, but not strong enough to create long-ranged orientational order, then the assumption of free rotors breaks down. Coupling and reduced symmetry means that all $3N$ modes are Raman active.

For strong interactions the molecular orientation is fixed and the modes are harmonic oscillations about that direction, known as librons.

This mixed character regime between a free rotor and harmonic oscillator is often described as "anharmonicity", which is trivially correct, in the sense that the modes are not harmonic. Unfortunately, "anharmonic" is often taken to mean that modes can simply be described with perturbation theory on a harmonic oscillator, and their selection rules remain largely unchanged. For strong interactions third-order anharmonic corrections to the librons may give a reasonable approximation, but as the amplitude of oscillation increases this perturbative approach will fail.

Even if the perturbation approach breaks down, we can retain the approximation that the modes are localised on a single molecule. The  wavefunctions remain localized even in the presence of strong interactions, provided the crystal has sufficient disorder \cite{ anderson1958absence, mendoza2016anderson}. This disorder can be in either molecular orientation, or in mass \cite{howie2014phonon, magduau2017infrared}.  Here we consider the calculation of Raman spectra in the limit of sufficiently strong disorder where the nuclear wavefunctions are localized on a single molecule with the surrounding molecules described by a mean-field potential. 

We will demonstrate that for a local potential with more than one inequivalent minimum, the eigenmodes include states which look like rotors and/or oscillators about different energy minima. The spectroscopic signal then contains rotor and oscillator modes, but also detrapping (oscillator $\rightarrow$ rotor) excitations, and reorientation modes (oscillator $\rightarrow$ oscillator) where the molecule moves between different minima pointing in different directions.
Rotor and oscillator modes have precise mass-dependencies where the transitions between energy levels scale with $1/m$ and $1/\sqrt{m}$, respectively. Reorientation modes are slightly more complicated because the transition energy comprises of two components: one related to the potential energy difference between the two minima, which is mass independent, and the other related to the zero point energy difference, which scales with $1/\sqrt{m}$. Therefore, the exact mass relationship of the reorientation modes depends on both the relative depth and curvature of the two potential minima: when the depths are very different there is no mass scaling.

For homo-atomic molecules mode parity leads to strong selection rules, which must be considered for the correct interpretation of the spectra.

We consider two diatomic rotors of mass $1$ and mass $2$ which we treat as proxies for hydrogen and deuterium respectively. Hydrogen forms a variety of solid phases, several of which are diatomic. The lowest pressure phase, phase I, consists of free rotors on a hexagonal lattice \cite{Silvera1980}. In phases II and III at higher pressure, the molecules exhibit preferred orientations but their centres of mass retain the hexagonal symmetry seen in phase I \cite{Silvera1981} \cite{hemley1988phase} \cite{goncharov2001spectroscopic}. The precise symmetry of these ordered phases is a topic of intense interest and many candidate structures have been proposed. \cite{Kohanoff1997} \cite{monserrat2016hexagonal} \cite{Pickard2007} \cite{azadi2017role} \cite{mcmahon2012properties}.

The parity of the rotor wavefunction also plays a significant role in the solid phases of hydrogen \cite{Eggert1999}. The overall symmetry of the wavefunction which is given by a combination of the nuclear and orientational wavefunction of the molecule, must remain antisymmetric (ferimions) or symmetric (bosons). Therefore for para-hydrogen, which has nuclear spin groundstate $\ket{\uparrow\downarrow-\downarrow\uparrow}$, the orientational wavefunction must have even parity. In the free rotor case this corresponds to states with $J=0,2,4...$. Similarly for ortho-hydrogen, where the nuclear spins are aligned, the orientational wavefunction must have odd parity, corresponding to states with $J=1,3,5...$ for the free rotor.

Several of the results we present have ramifications for interpretation of the Raman spectra of phases of hydrogen. However the potential operators applied here are far simpler than those present in hydrogen and deuterium solids and we therefore refrain from direct comparison with experiment.

\section{Methods}
\label{sec:Methods}

\subsection{Rotor, Oscillator and Reorientational Modes}
    
 The Raman active transitions of diatomic molecules are generally modelled as two distinct types of excitation; rotational excitations or 'rotons' and vibrational excitations, either 'librons' or 'vibrons'.
 
 Rotational excitations are well described for a diatomic rotor; the roton energy levels are given in 2D by:
\begin{equation} E(J) = \frac{\hbar^2}{mr^2}J^2 \label{rotonenergy2D} \end{equation} and in 3D by:
\begin{equation} E(J) = \frac{\hbar^2}{mr^2}J\left(J+1\right) \label{rotonenergy} \end{equation} where $r$ is the bond length, 
$m$ is the atomic mass and $J$ is an integer quantum number. The selection rules for Raman transitions for the free rotor are $\Delta J = 0, \pm 2$.

Vibrons and librons are described as quantum harmonic oscillators for which energy levels are given by:
\begin{equation} E(n) = \hbar\omega\left(n+\frac{1}{2}\right) =\sqrt{ \frac{\hbar^2k}{m}} 
\left(n+\frac{1}{2}\right) \end{equation} 
with $\omega$ the frequency and $k$ the effective spring constant. 

The different energy-mass relationships for the free rotor ($1/m$) and harmonic oscillator ($1/\sqrt{m})$ provide a way to characterise Raman modes using different isotopes of a diatomic molecule.\cite{cookeraman} \cite{pena2020quantitative}
Here, we will compare the frequencies of two specific molecules, with atomic mass 1 (``hydrogen'') and 2 (``deuterium''):

\begin{equation} R =  \frac{E_i(H)-E_j(H)}{E_i(D)-E_j(D)} = \upsilon_{H_2}/\upsilon_{D_2}
\label{eq:R_definition}
\end{equation}

\noindent  we expect a rotor to give an isotopic energy ratio $R=2$ and an oscillator to give $R=\sqrt{2}$, where $E_i$ and $E_j$ are the calculated energy levels and $\upsilon$ represents the experimentally-measurable Raman shift. For a reorientional mode we expect a weak mass dependence. These modes give rise to a transition from an oscillator state in one minimum in e.g. the y-direction, to a new oscillator state in an inequivalent minimum in e.g. the z-direction. We therefore expect the Raman shift for this mode to be given by:

\begin{equation}
\nu=\frac{1}{\sqrt{m}}(\sqrt{k_{y}}-\sqrt{k_{z}})+\Delta E_{min}
\end{equation}

and the value of R to be given by:
\begin{equation}
    R=\sqrt{2}\frac{1+(\sqrt{k_{y}}-\sqrt{k_{z}})/\Delta E_{min}}{\sqrt{2}+(\sqrt{k_{y}}-\sqrt{k_{z}})/\Delta E_{min}}
\end{equation}

where $\Delta E_{min}$ is the difference in energy between the two minima. If $\Delta E_{min}$ is large with respect to $\sqrt{k_{y}}-\sqrt{k_{z}}$ we expect a value of R close to $1$. If $\Delta E_{min}$ is small with respect to $\sqrt{k_{y}}-\sqrt{k_{z}}$ we expect $R\approx\sqrt{2}$. 

Whilst the free rotor or oscillator models can describe the majority of excitations in diatomic systems, there are certain solids for which this approach is insufficient. In particular solid phases of hydrogen where nuclear quantum effects are significant. Here we describe a model for a system that sits between the idealised cases of rotors and oscillators:  the hindered rotor.

\subsection{Modelling a hindered rotor}

We consider only the angular dependent modes: rotons and librons. Starting with the angular Schroedinger equation:

\begin{equation}
\begin{aligned}
H(\theta,\phi) = &- \frac{\hbar^2}{mr^2} \left [\frac{1}{\sin \theta} \frac{\partial}{\partial \theta} \left ( \sin \theta \frac{\partial}{ \partial \theta} \right ) + \frac{1}{{\sin^2 \theta}} \frac{\partial^2}{\partial \phi^2} \right]
\\&+ V(\theta, \phi) \label{eq:3D}
\end{aligned}
\end{equation}

where $r=\mathbf{|r|}$ is the molecular bond length and $m$ is the mass of the nucleus.

We introduce a simple mean-field potential to describe the angular dependence due to local crystal structure.  This is conveniently expanded in spherical harmonics.

\begin{equation}
    V(\theta, \phi) = \sum_{l'',m''}{c_{l'',m''}Y_{l'',m''}(\theta,\phi)}
\end{equation}

giving the Hamiltonian in the basis of the free rotor as:

\begin{equation}
H_{lml'm'}^{(0)} = \frac{\hbar^2}{2{\mu}r^2}l(l+1)\delta_{ll'} \delta_{mm'} + Y_{lm}V(\theta,\phi)Y_{l',m'}
\end{equation}

where the primes distinguish spherical harmonics used for the potential and wavefunctions.

We then diagonalise the Hamiltonian to find the energy levels $n$:

\begin{equation}
H_{nn'}^{(0)} = D^{*}_{n,lm} H_{lml'm'}^{(0)} D_{l'm',n'}
\end{equation}

with hindered-rotor eigenstates now given by:

\begin{equation}
\psi_{n} = D^{*}_{n,lm} Y_{lm}
\end{equation}

These states are associated with the appropriate nuclear spin isomer to form overall antisymmetric (fermionic) or symmetric (bosonic) states.  The nuclear wavefunction does not affect the eigenstate energy, or the polarizability tensor, but it does affect the degeneracy and therefore the intensity of the Raman signal. For H$_2$ there are singlet and triplet nuclear states, D$_2$ has singlet, triplet and quintuplet.

\subsection{Calculating the Raman signal}

To evaluate the transitions between states we express the polarizability operator in the hindered rotor basis, this is done through a unitary transformation on the polarizability tensor of the free molecule. We then calculate the  Raman signal using the response of the quantum system to a sudden excitation. We compute the time-dependent response to excitation using the Liouville-von Neumann (LvN) equation, and then evaluate the expectation value of the resulting polarization \cite{cookeraman, mukamel1995principles, magdau2019interpretation, mead2020sum}. The Raman spectrum is then obtained by Fourier transform (FT) of the response in the time domain. Full details are given in our previous work \cite{cookeraman} and references within. This approach automatically satisfies selection rules based on transition polarizability and nuclear exchange parity.

\section{Results and Discussion}
\label{sec:Results}

\begin{figure}[H]
\begin{center}
\includegraphics[width=0.8\linewidth]{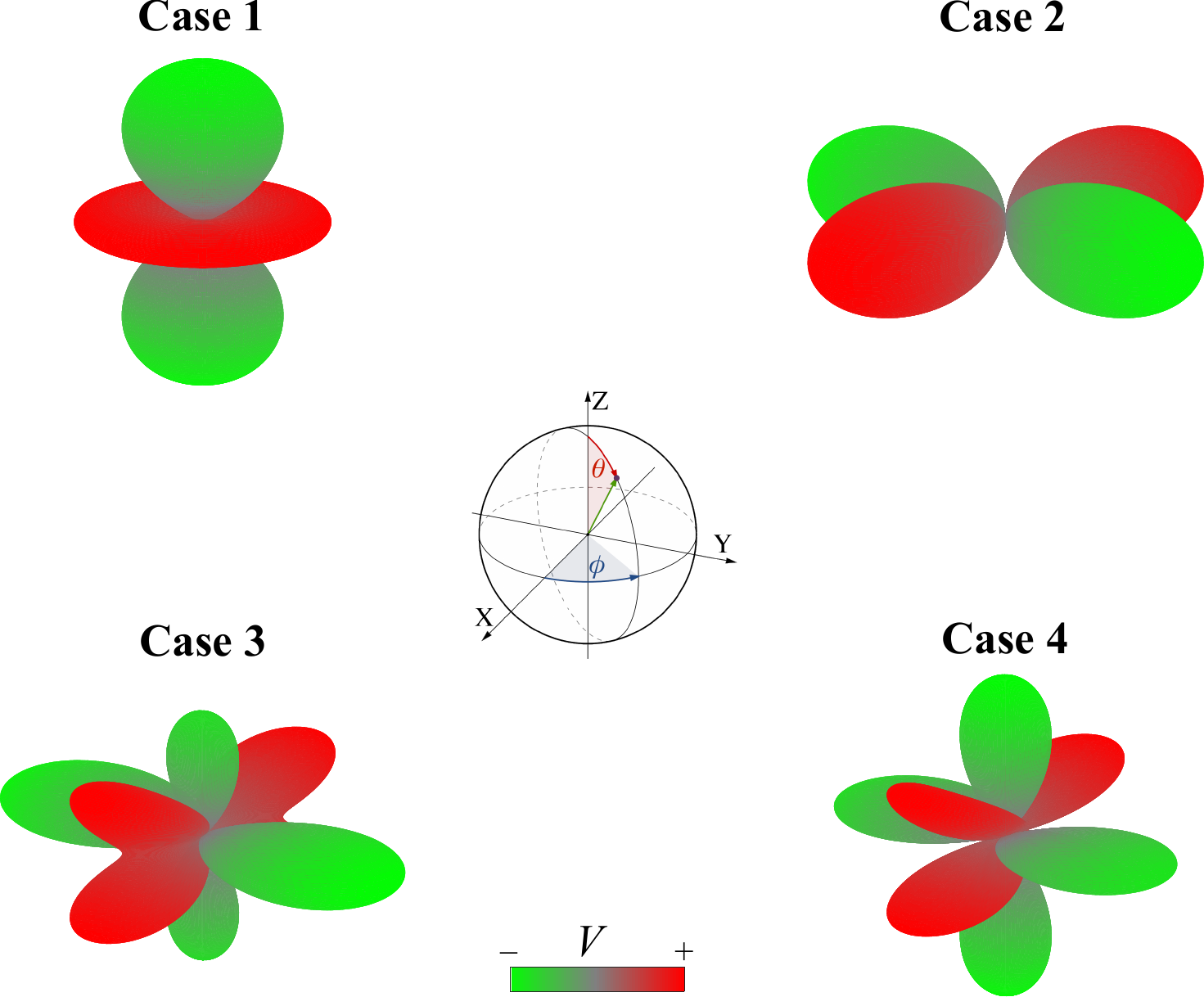}
\end{center}
\caption{Potential isosurfaces for the four cases considered.  Distance of the surface from the origin represents the strength of the potential, green is attractive, red repulsive. Case 1 has a potential minimum for the molecule pointing in the z-direction. Case 2 favours the y-orientation, case 3 has two minima with different depths in y- and z-directions, case 4 has two minima with similar depth.}
\label{fig:Potentials}
\end{figure}

We present results for four different  potential energy surfaces (Fig \ref{fig:Potentials}), gradually building up in complexity.  In each case, the weak field limit is a perturbed rotor, while our model also explores the strong field regime.

The simplest, case 1, is a potential with a  single minimum along z-direction with equal curvature in both x- and y-directions; in the limit of strong field we expect to find a single oscillator mode degenerate along x and y. In case $2$ we introduce a potential  with a minimum along y where the curvature is different in spatial dimensions x and z: here we expect to see an asymmetric 2D oscillator with modes with different frequencies in the two dimensions. Case $3$ represents a combination of the previous two cases, there are now $2$ inequivalent minima,  parallel to the y and z axes. In case $4$ we change the relative depth of these inequivalent minima.

All of these model potentials have higher symmetry than the hcp structure of solid hydrogen phase I, so they will have simpler spectroscopic signatures.

\subsection{Case 1: $V=PY_{20}$}

We start with a simple potential $V=PY_{20}$ that only has variation in $\theta$ and is symmetric in $\phi$. In this potential, the lowest energy state has the molecule pointing along the z-direction.  To first order in perturbation theory, the potential is quadratic and symmetric. The scale of the potential, P, is arbitrary, but it is understood that it will vary monotonically with pressure.

\begin{figure}[H]
\begin{center}
\includegraphics[width=0.6\textwidth]{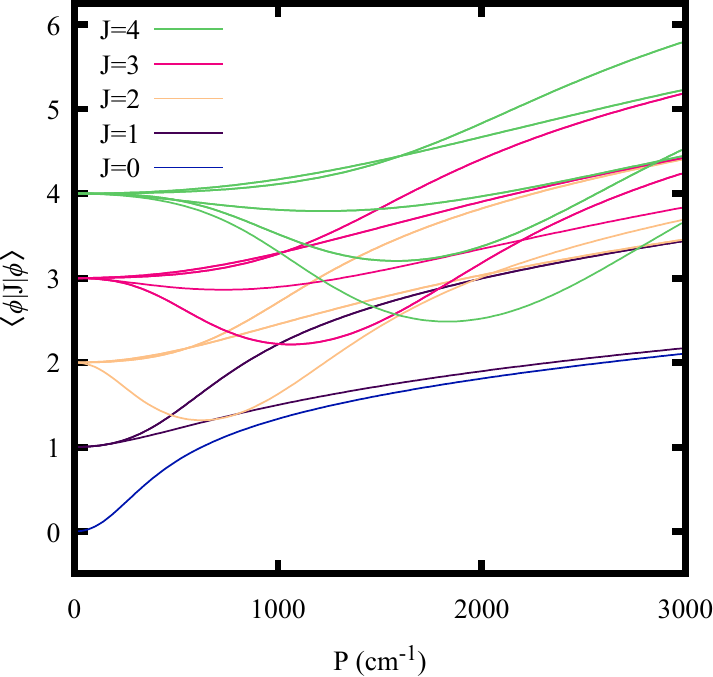}
\end{center}
\caption{Expectation value of J, $\bra{\phi}\hat{J}\ket{\phi}$ for the case 1 potential. Modes are colored according to their $J$ values at $P=0$. By $P=1000\:\mathrm{cm^{-1}}$ there is significant mixing between the free rotor basis states. However, the rotational symmetry of the potential around the z-axis restricts mixing to states with equal $|m_J|$.}
\label{fig:Jvalue}
\end{figure}

\begin{figure}[H]
\begin{center}
\includegraphics[width=0.8\textwidth]{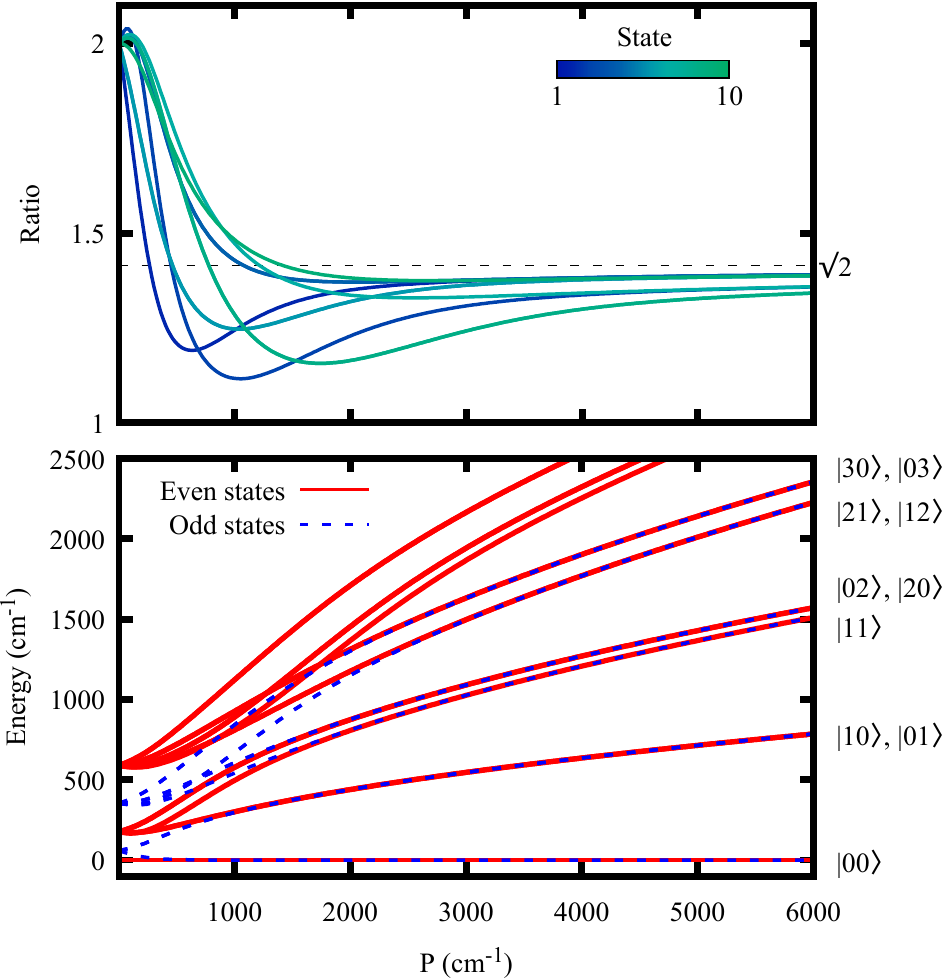}
\end{center}
\caption{Bottom: energy levels for the first twenty-five states of deuterium for case 1  ($V = PY_{20}$) relative to the ground state. Energies units are expressed in cm$^{-1}$ by setting mass and bond-length to correspond to the deuterium molecule. States with the same absolute value of $m_J$ remain degenerate because the potential is rotationally symmetric around z. As a result, each original state with $2J+1$ degeneracy appears to split into $J+1$ separate states. Top: isotope energy ratios R for case 1 for first ten excited states against the potential strength P (see eq. \ref{eq:R_definition} for definition). As P is increased the ratio R transitions from $2$ for a free rotor to $\sqrt{2}$ (dashed line) for a harmonic oscillator.}
\label{fig:Energy1}
\end{figure}

\begin{figure}[H]
\begin{center}
\includegraphics[width=1.0\linewidth]{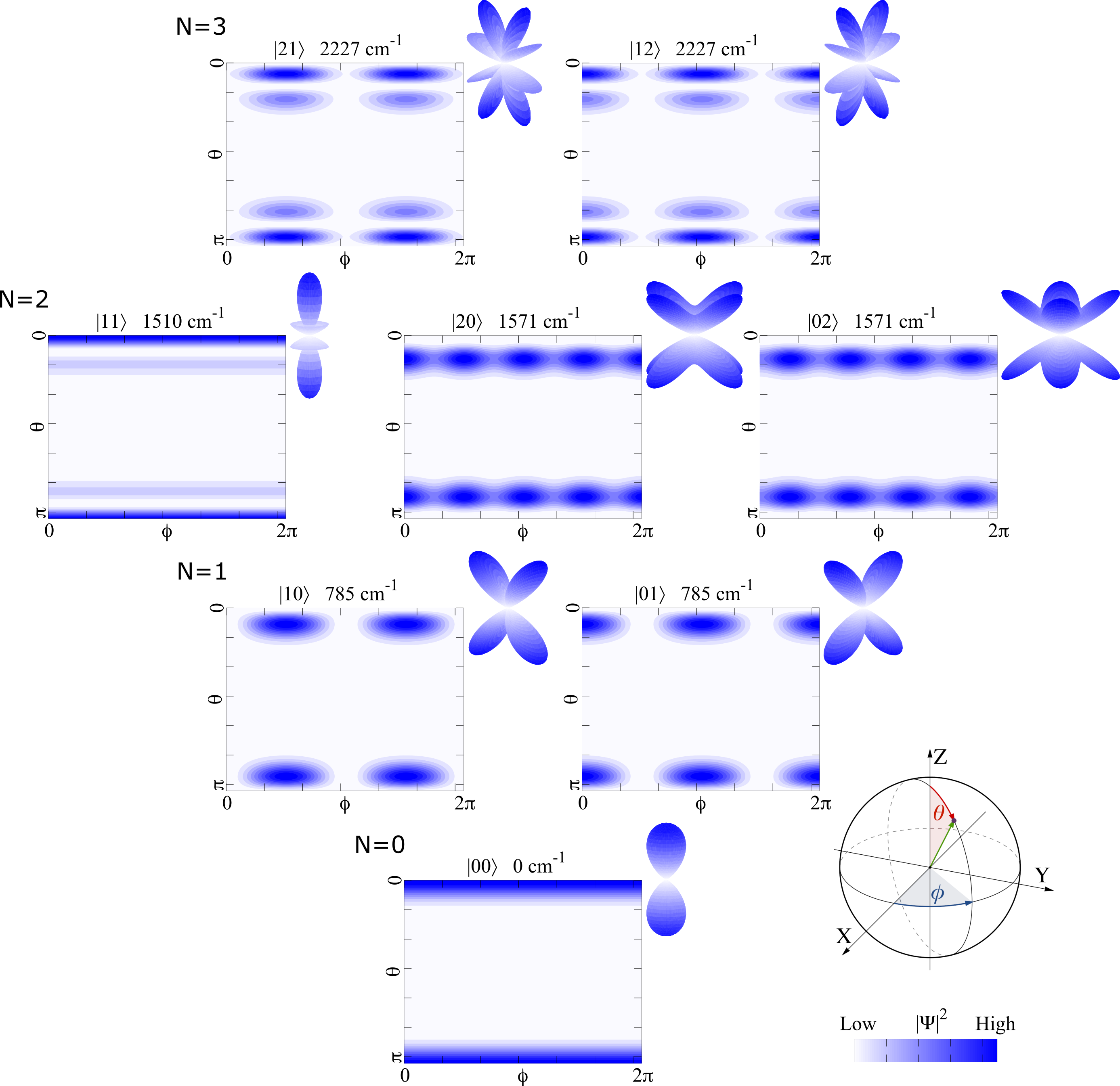}
\end{center}
\caption{First $8$ odd-parity wavefunctions of deuterium obtained at $P=6000\hspace{5pt} \mathrm{cm^{-1}}$ for case 1, shown with contour plots in ($\theta,\phi$) and 3D isosurfaces of $|\Psi|^2$. The subplots are arranged based on the total number of quanta $N$ in each state.}
\label{fig:wfn1}
\end{figure}

The energy levels as a function of potential strength P are shown in Fig. \ref{fig:Energy1} with the corresponding wavefunctions in Fig. \ref{fig:wfn1}. At zero potential, we recover the free rotor states, which can be well described by quantum numbers $|J,m_J\rangle$ and have degeneracies 1, 3, 5 ... (2J+1) with increasing energy. Figure \ref{fig:Jvalue} shows that in intermediate field, $J$ is no longer a good quantum number and all states acquire mixed character. However in case 1, mixing in $m_J$ is restricted to basis functions with equal $|m_J|$ due to the rotational symmetry of the potential around the z-axis.  All states can be assigned a clear parity and consist of entirely odd or entirely even $J$ basis functions. States of opposite parity pair up  with asymptotically degenerate energies towards the strong field (pure oscillator regime) as shown in Fig. \ref{fig:Energy1}.

In the limit of high potential the molecule is oriented along z ($\theta=0$ in polar coordinates), and its excitations can be described using eigenstates of a 2D harmonic oscillator $|n_1,n_2\rangle$. The ground state is $|0,0\rangle$ - a doublet between even and odd parity.  The first excited states are degenerate $|1,0\rangle$, $|0,1\rangle$, a quadruplet when accounting for parity degeneracy. The second group of six excited states are $|2,0\rangle$, $|0,2\rangle$, and $|1,1\rangle$. For a symmetric harmonic oscillator these are still degenerate, but because the potential is anharmonic the $|1,1\rangle$ state has slightly lower energy. The third group of eight states: $|3,0\rangle$, $|0,3\rangle$, $|2,1\rangle$ and $|1,2\rangle$, comprises two quadruplets.

Spectroscopy cannot measure the eigenstate energies, rather it measures energy differences between eigenstates: the Raman shifts relate to selected ($E_i-E_j$) differences. As noted above, the energies depend on isotope mass, and in Fig. \ref{fig:Energy1} we show the isotope energy ratio for transitions $E_0 \rightarrow E_i$ for the first ten excited states $i$, in otherwise-identical molecules with atomic masses 1 and 2, respectively. Values of R move smoothly from $2\rightarrow\sqrt{2}$, for all transitions shown, some of which are Raman active.  Because the potential is softer than harmonic at large amplitudes, R overshoots and approaches $\sqrt{2}$ from below.

\begin{figure}[H]
\begin{center} 
\includegraphics[width=1\linewidth]{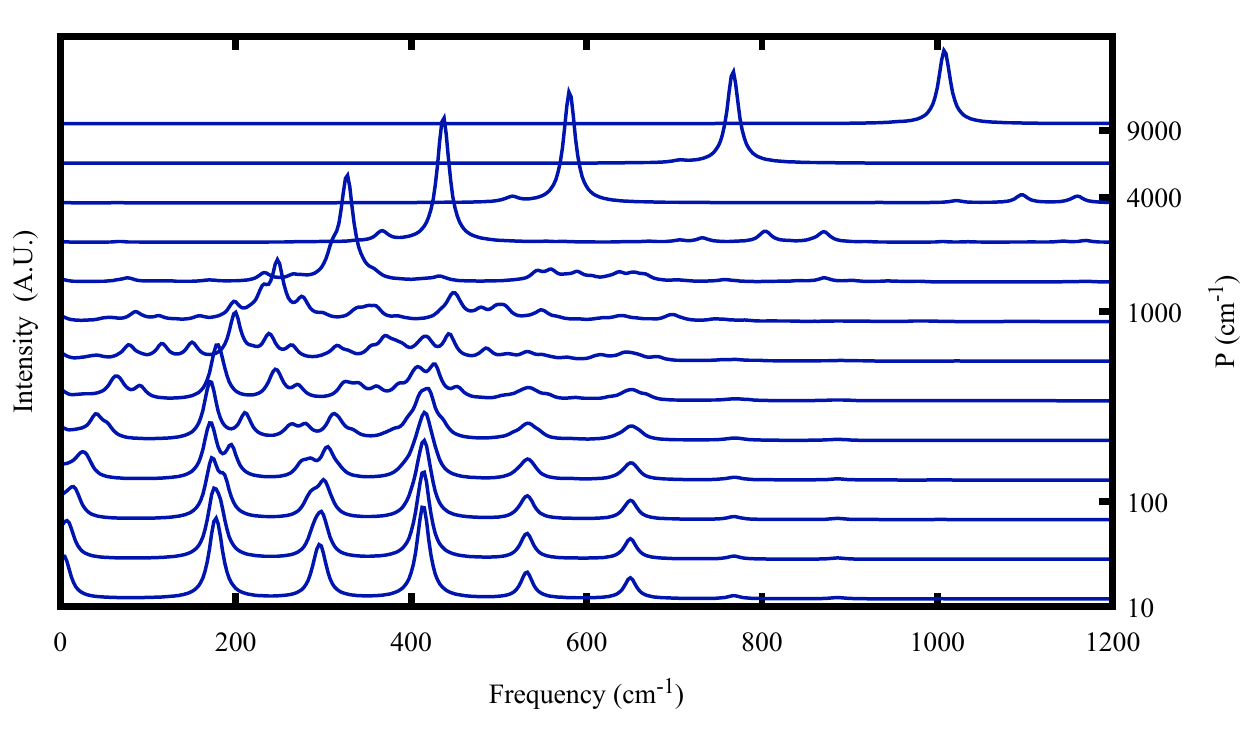}
\end{center}
\caption{Calculated Raman spectra vs P for deuterium in case 1. The right hand y-axis indicates the value of P on a logarithmic scale. We assume a temperature of 300 K. At low values of P the spectra resemble the free rotor regime with a series of $\Delta J=2$ roton transitions visible. At high values of P a single peak is visible corresponding to an oscillator transition ($\ket{0,0} \rightarrow \ket{0,1}$ shown in Fig. \ref{fig:wfn1}).}
\label{fig:Raman1}
\end{figure}

Figure \ref{fig:Raman1} shows the calculated Raman spectrum for the simplest case 1. At low potential we see the familiar form of the quantum rotors, and at high potential the single peak of a harmonic oscillator (libron). The intermediate potential region is characterised by a very complicated spectrum and the appearance of many new peaks, despite the symmetry of the potential remaining unchanged.  These arise from both splitting of the rotor peaks, and from transitions which acquire Raman activity due to the symmetry-breaking potential lifting the rotor selection rules.

\subsection{Case 2: $V=P(Y_{22} + Y_{2-2})$}
\begin{figure}[H]
\begin{center}
\includegraphics[width=1.0\linewidth]{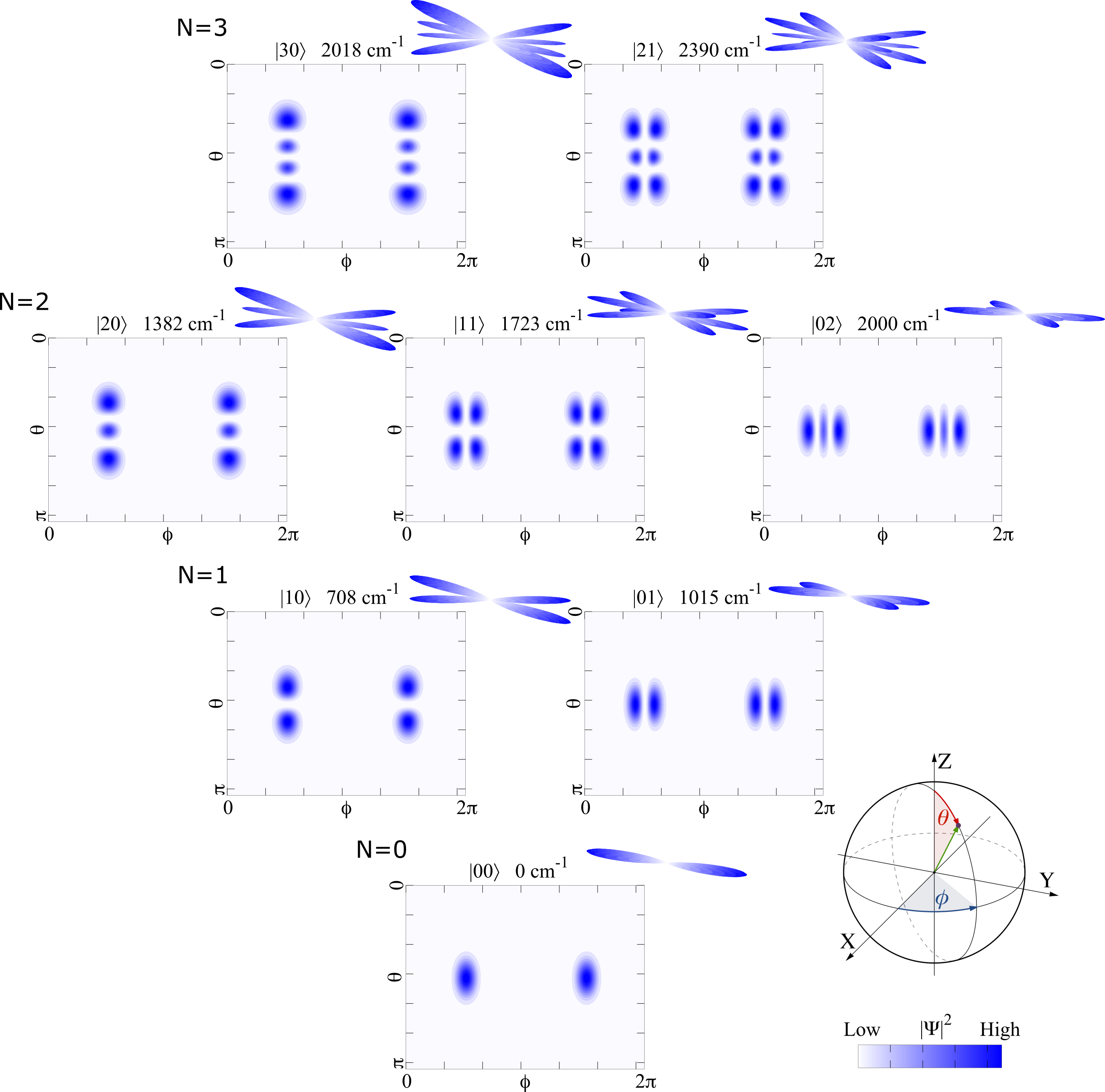}
\end{center}
\caption{First $8$ even-parity wavefunctions of deuterium obtained at $P=6000 \hspace{5pt} \mathrm{cm^{-1}}$ for case 2, shown with contour plots in ($\theta,\phi$) and 3D isosurfaces of $|\Psi|^2$. The subplots are arranged based on the total number of quanta $N$ in each state. States with the same number of quanta (e.g. $\ket{01}$ and $\ket{10}$) are no longer degenerate due the difference in curvature in $\theta$ and $\phi$ of the potential minimum.}
\label{fig:wfn2}
\end{figure}

Next we consider a potential of the form $V=P(Y_{22} + Y_{2-2})$ that has a single minimum pointing in the y-direction and with different curvature in $\theta$ and $\phi$. This has a ground state with the molecule pointing along the potential minimum (Fig. \ref{fig:wfn2}). The reduced symmetry of this potential lifts all $m_J$ degeneracies in the rotor states.

\begin{figure}[H]
\begin{center}
\includegraphics[width=0.7\linewidth]{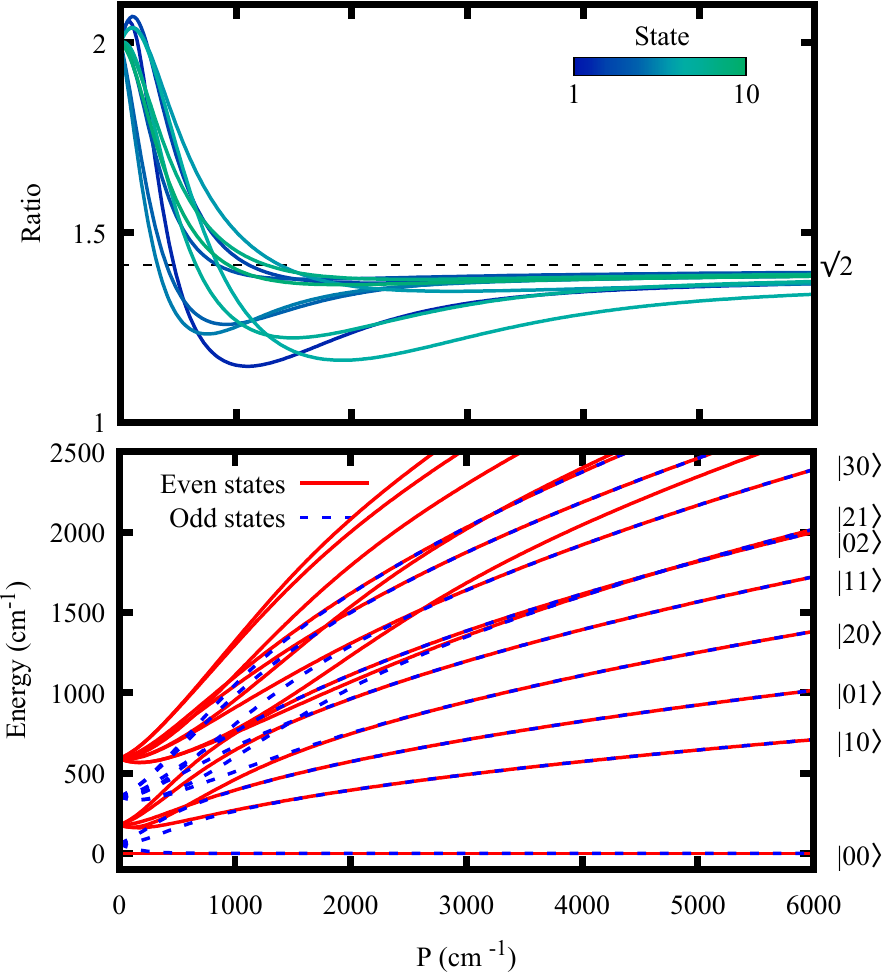}
\end{center}
\caption{Bottom: energy levels for the first twenty-five states of deuterium for case 2  ($V = P(Y_{22} + Y_{2-2})$) relative to the ground state. The degeneracy in $|m_J|$ seen in case 1 is now broken by the reduced symmetry of the case 2 potential. This lower symmetry also lowers the degeneracy of the oscillator states in the strong field limit, from four-fold (seen in case 1) to two-fold. Top: isotope energy ratios (R) for first ten excited states in case 2 against the potential strength P. As with case 1, R transitions from $2$ for a free rotor to $\sqrt{2}$ (dashed line) for a harmonic oscillator.}
\label{fig:Energy2}
\end{figure}

As with case 1, in the strong potential limit there are degenerate odd and even parity states which can be labelled with 2D SHO quantum numbers $|n_1,n_2\rangle$, in energy order: $|0,0\rangle$, $|1,0\rangle$, $|0,1\rangle$, $|2,0\rangle$, $|1,1\rangle$, $|0,2\rangle$, $|3,0\rangle$ and $|2,1\rangle$ (Fig. \ref{fig:Energy2}, \ref{fig:wfn2}). In this limit, the energy levels all become even-odd parity pairs, and energies approximate to $(n_1+\frac{1}{2})\hbar\omega_1 + (n_2+\frac{1}{2})\hbar\omega_2$.

\begin{figure}[H]
\includegraphics[width=1.0\linewidth]{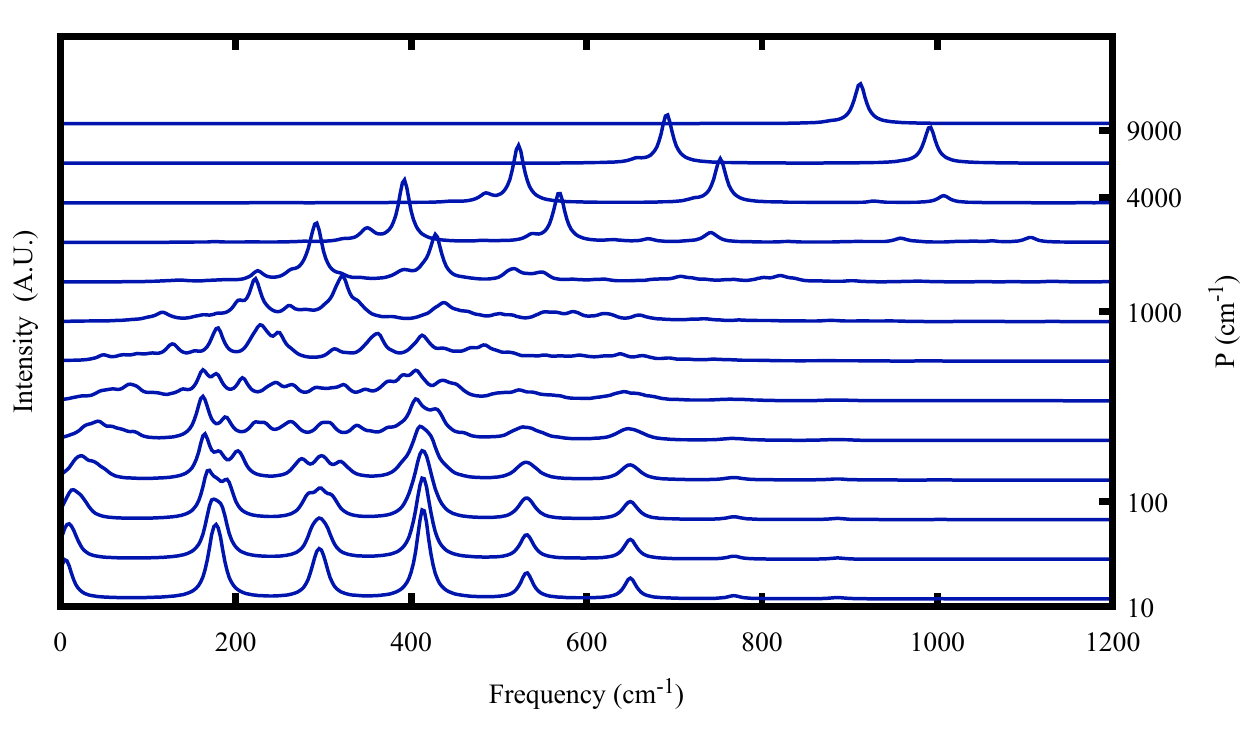}
\caption{Raman spectra for case 2 with deuterium. Right-hand y-axis indicates the offset of the spectra by the value of P on a log scale. As with case 1 (Fig. \ref{fig:Raman1}), the spectrum undergoes a dramatic transformation from rotor states to libration states upon increasing pressure. Unlike in case 1, the 2D SHO degeneracy is lifted by the different curvatures in $\theta$ and $\phi$, resulting in two distinct libron modes at high potential strength.
}
\label{fig:Raman2}
\end{figure}

\subsection{Case 3. $V=P(Y_{22} + Y_{2-2} - Y_{40})$}
\begin{figure}[H]
\begin{center}
 \includegraphics[width=0.7\linewidth]{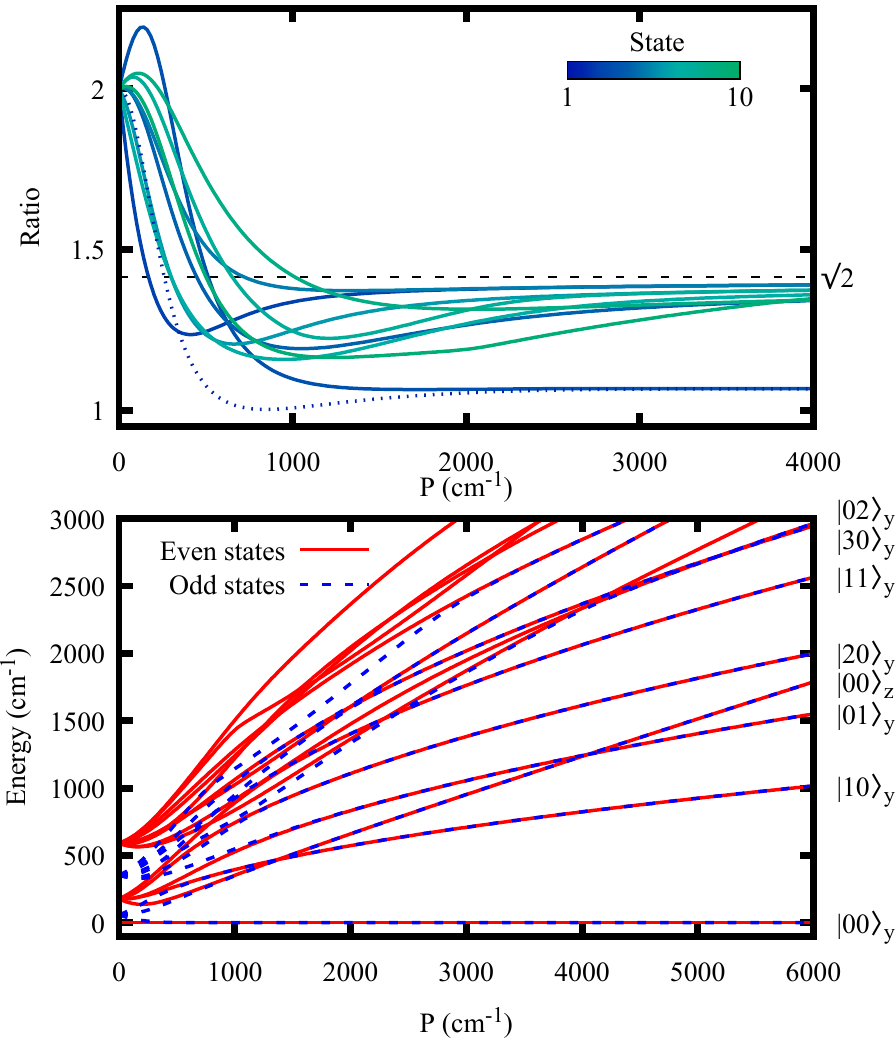}
 \end{center}
 \caption{Bottom: Energy levels for first twenty-five states of deuterium for case 3 ($V = P( Y_{22} + Y_{2-2} - Y_{40})$) relative to the ground state. A metastable state exists corresponding to orientation along the z-axis ($\ket{00}_z$ in Fig. \ref{fig:wfn3}). The system now consists of two sets of independent oscillator states, one for each minimum.  Top: isotope energy ratios (R) for first ten excited states for case 3 against the potential strength P. The value of R now transitions from $2$ for a free rotor to either $1$ or $\sqrt{2}$ for a re-orientation or harmonic oscillator, respectively. The dashed line indicates the re-orientational transition corresponding to the $\ket{00}_y \: \rightarrow \: \ket{00}_z$ transition shown in Fig. \ref{fig:ReorientRaman1}.}
 \label{fig:Energy3}
\end{figure}

\begin{figure}[H]
\begin{center}
\includegraphics[width=1.0\linewidth]{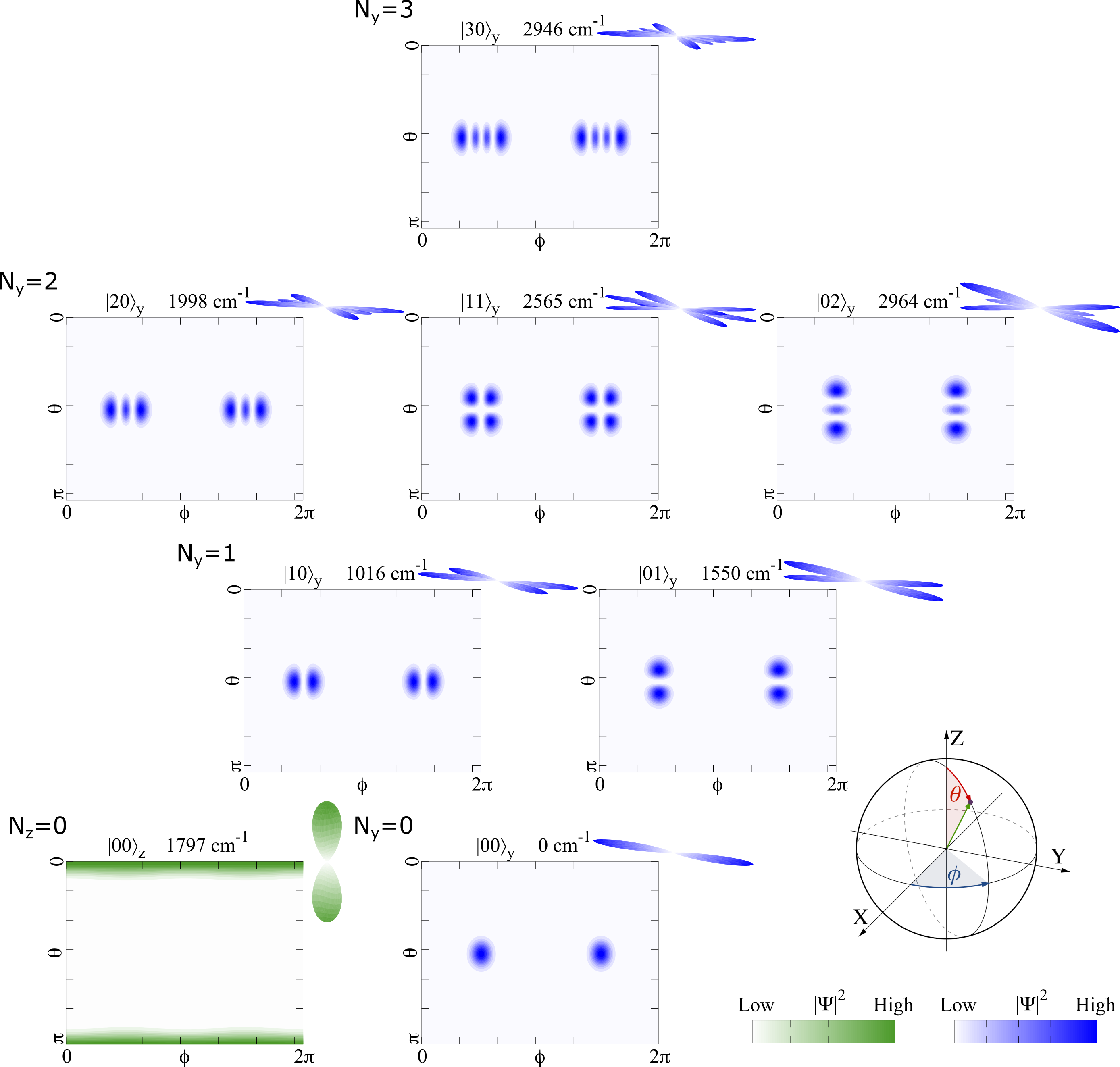}
\end{center}
\caption{First $8$ even-parity wavefunctions of deuterium obtained at $P=6000 \hspace{5pt} \mathrm{cm^{-1}}$ for the case 3 potential, shown with contour plots in ($\theta,\phi$) and 3D isosurfaces of $|\Psi|^2$. The subplots are arranged based on the total number of quanta $N$ in each state. The two different oscillators along z and y are shown in different colors, green and blue, respectively. }
\label{fig:wfn3}
\end{figure}

\begin{figure}[H]
\begin{center}
\includegraphics{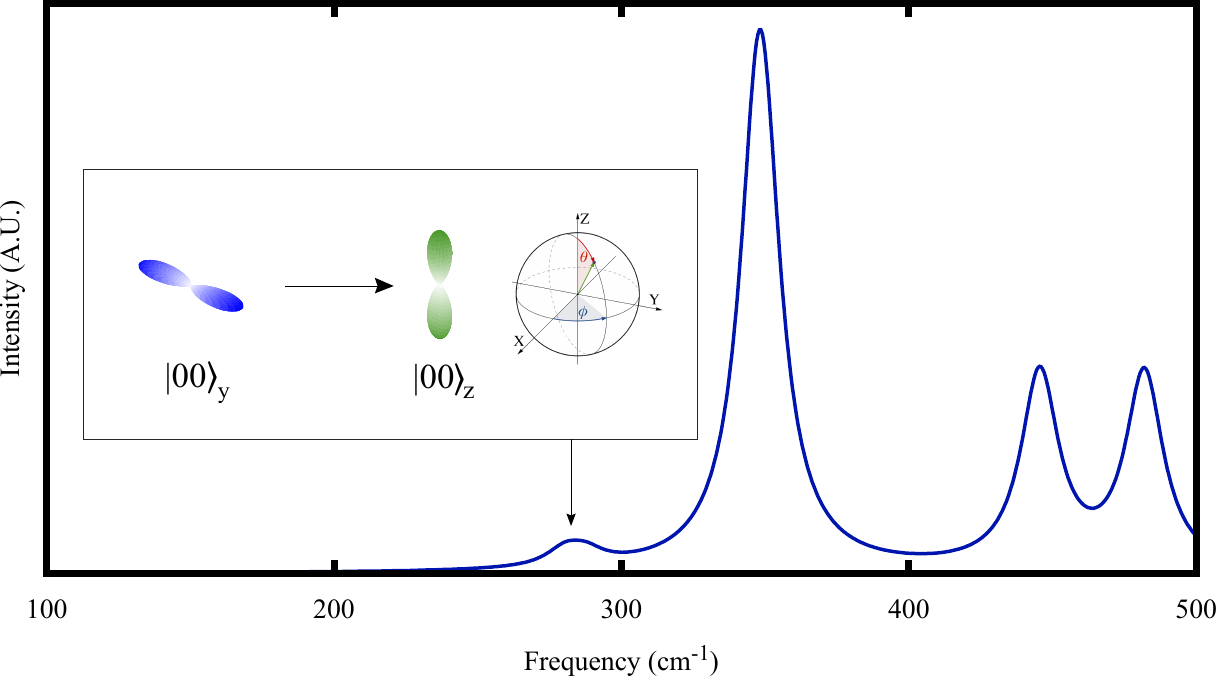}
\end{center}
\caption{Raman Spectrum at P$=800$ cm$^{-1}$ with reorientaional mode labelled. The temperature in this case is set to 10 K to allow clear identification of the re-orientational mode.}
\label{fig:ReorientRaman1}
\end{figure}

The third case is a potential with two distinct minima, along the z- and y-directions.  In this case the classical energy minimum has the molecule pointing along the y-axis, with a metastable minimum for molecules pointing along z.  Although this is still simpler than any real system, we can see in Fig. \ref{fig:Energy3} that for intermediate strength potential the energy levels have no clear structure.  The high energy limit still tends to be harmonic oscillator-like, and is sensitive to the relative stability of the two minima.  In this case the global minimum is deep enough that the lowest three doublet levels are all related to it and can be labelled $|n_1,n_2\rangle_y$: $|0,0\rangle_y$, $|1,0\rangle_y$ and $|0,1\rangle_y$.  Fig. \ref{fig:wfn3} shows that the metastable state, pointing in the z-direction, has an energy of $~1800 \: \mathrm{cm^{-1}}$ higher than the ground state. This state can be described with two further quantum numbers $|n_1',n_2'\rangle_z$ shown as $|0,0\rangle_{z}$ in Fig. \ref{fig:wfn3}. In the energy plot (Fig. \ref{fig:Energy3}) we observe that the energy of this state relative to the global ground state rises approximately linearly with the potential, starting from $J=1$ and $J=2$ states in the free rotor regime. Fig. \ref{fig:ReorientRaman1} shows that the $\ket{00}_y \rightarrow \ket{00}_z$ transition is Raman active  and corresponds to the re-orientation of the molecule from one minimum to another. For this model potential, the re-orientation can be observed at low temperature and intermediate field strength. The isotopic mass ratio for this transition is close to 1 because the transition energy is dominated by the potential energy difference between the two minima.

\begin{figure}[H]
\includegraphics[width=1.0\linewidth]{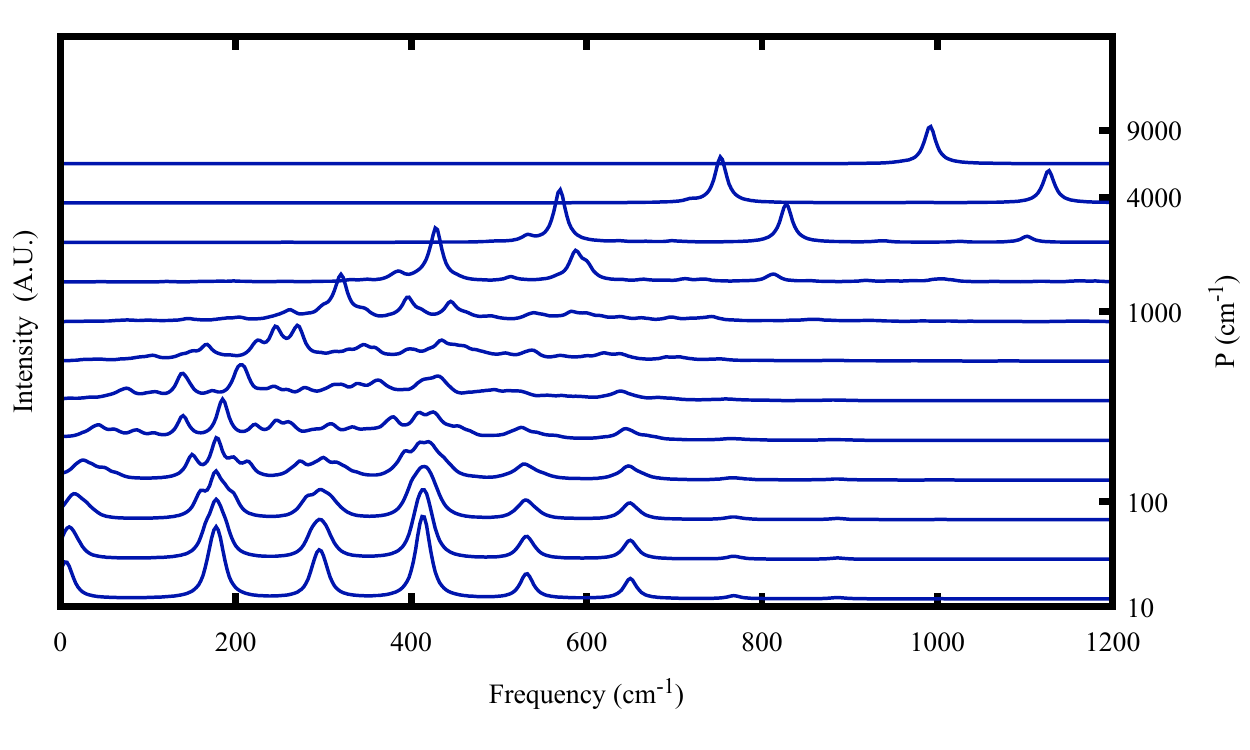}
\caption{Raman spectra for case 3 with deuterium at 300 K. Right-hand y-axis indicates the offset of the spectra by the value of P on a log scale. Despite the system being composed of two independent oscillators, the Raman spectra broadly resemble case 2, with a single oscillator with two libron modes at high field. This is because the metastable $\ket{00}_z$ state is not populated at 300 K.}
\label{fig:Raman3}
\end{figure}

\subsection{Case 4. $V=P(Y_{22} + Y_{2-2} - 2Y_{40})$}
\begin{figure}[H]
\begin{center}
\includegraphics[width=1.0\linewidth]{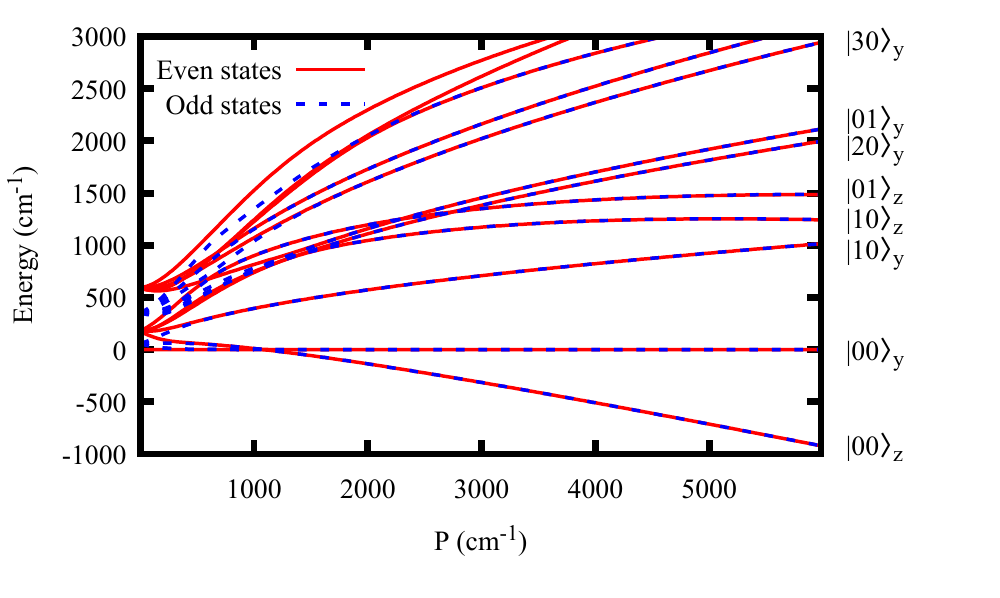}
\end{center}
\caption{Energy levels relative to the reference state $\ket{00}_y$ for D$_2$  where $V = P ( Y_{22}+Y_{2-2} - 2Y_{40} )$. In this case the two minima in the potential are of similar depth. The combined effect of slightly differing curvatures and depths of the two minima leads to a crossover of the ground state at $ P \approx 1000\: \mathrm{cm^{-1}}$}
\label{fig:Energy4}
\end{figure}

The final case also involves a potential with minima along y and z,  but here the two orientations are close in energy. There is an unusual situation where, as the potential increases,  the ground state switches its character from pointing along y to pointing along z.  To illustrate the crossover we plot the eigenstate energies in Fig. \ref{fig:Energy4}, relative to the reference state $\ket{00}_y$ rather than the difference from the ground state as in the previous figures.  The crossover can be understood as a 
trade-off between the zero point energies and the potential energy of the two minima pointing in the y- and z-directions. The state pointing along y has a higher potential energy minimum, but it is stabilised by a lower zero point energy. As the potential energy difference between the two minima increases with field strength, while the zero point energy remains unchanged, the state pointing in the z-direction eventually becomes more stable and hence a crossover takes place at around $P=1000\:\mathrm{cm}^{-1}$.

\begin{figure}[H]
\begin{center}
\includegraphics[width=1.0\linewidth]{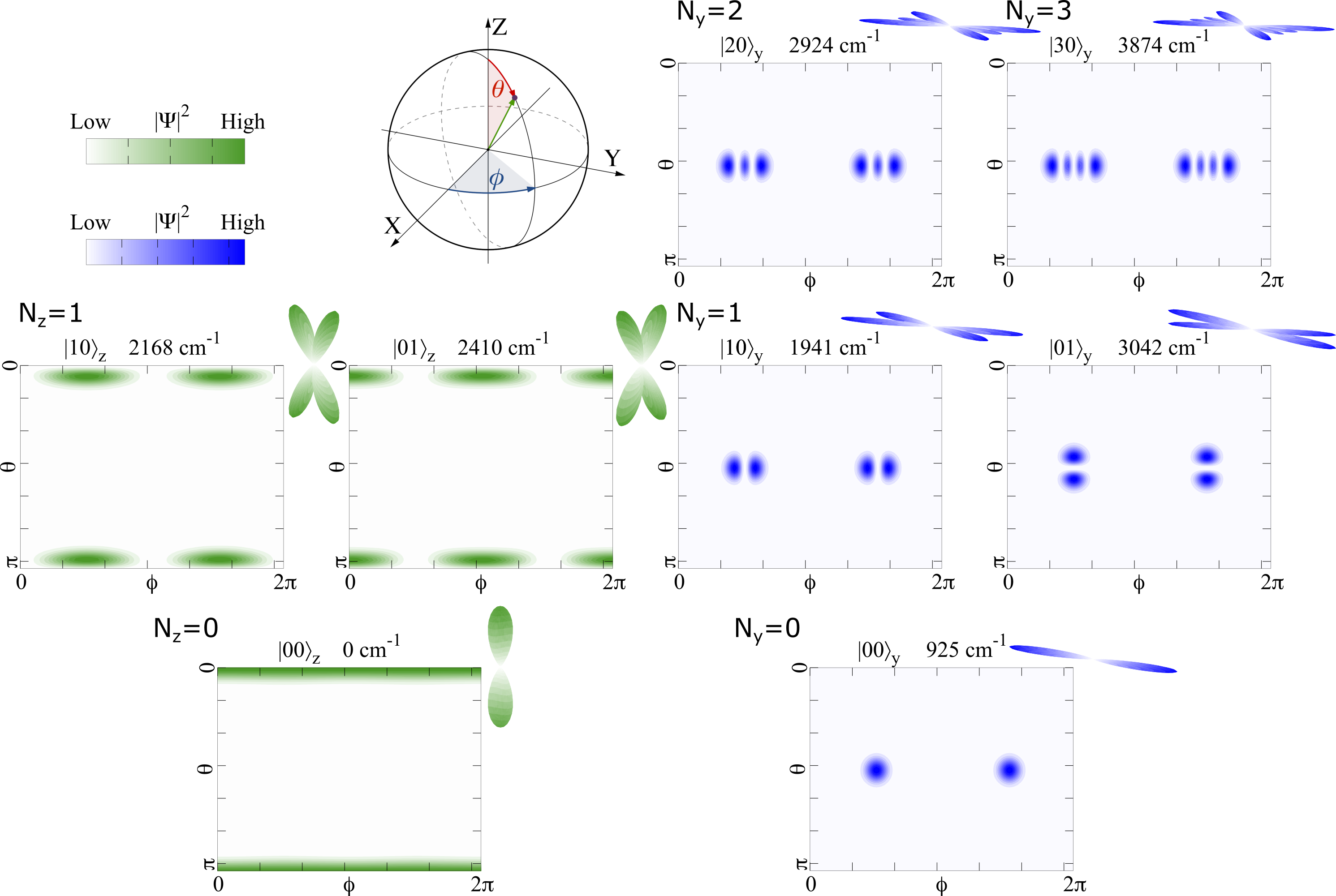}
\end{center}
\caption{First $8$ even-parity wavefunctions of $D_2$ obtained at $P=6000 \hspace{5pt} \mathrm{cm^{-1}}$ for the case 4 potential, shown with contour plots in ($\theta,\phi$) and 3D isosurfaces of $|\Psi|^2$. The subplots are arranged based on the total number of quantas $N$ in each state. The two different oscillators in z and y are shown in different colors, green and blue, respectively. Energies are given relative to the global minimum ($\ket{00}_z$) state.}
\label{fig:wfn4}
\end{figure}

\begin{figure}[H]
\includegraphics[width=1.0\linewidth]{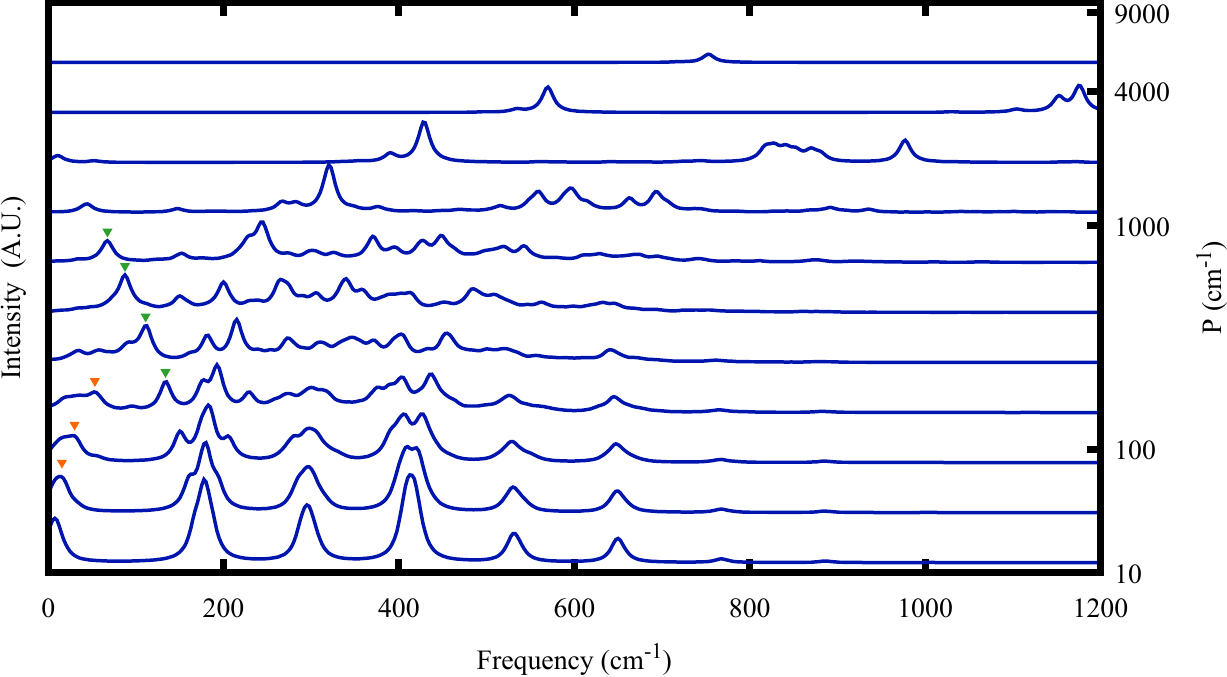}
\caption{Raman spectra for case 4 with $D_2$ at 300 K. Right-hand y-axis indicates the value of P on a log scale. The two independent oscillators are closer in energy in this case and consequently the spectra at high field is composed of contributions from each system. Indicated with triangles are two kinds of re-orientational modes: rotational axis re-orientation (orange) and directional re-orientation (green).} 
\label{fig:Raman4}
\end{figure}

\begin{figure}[H]
\begin{center}
\includegraphics{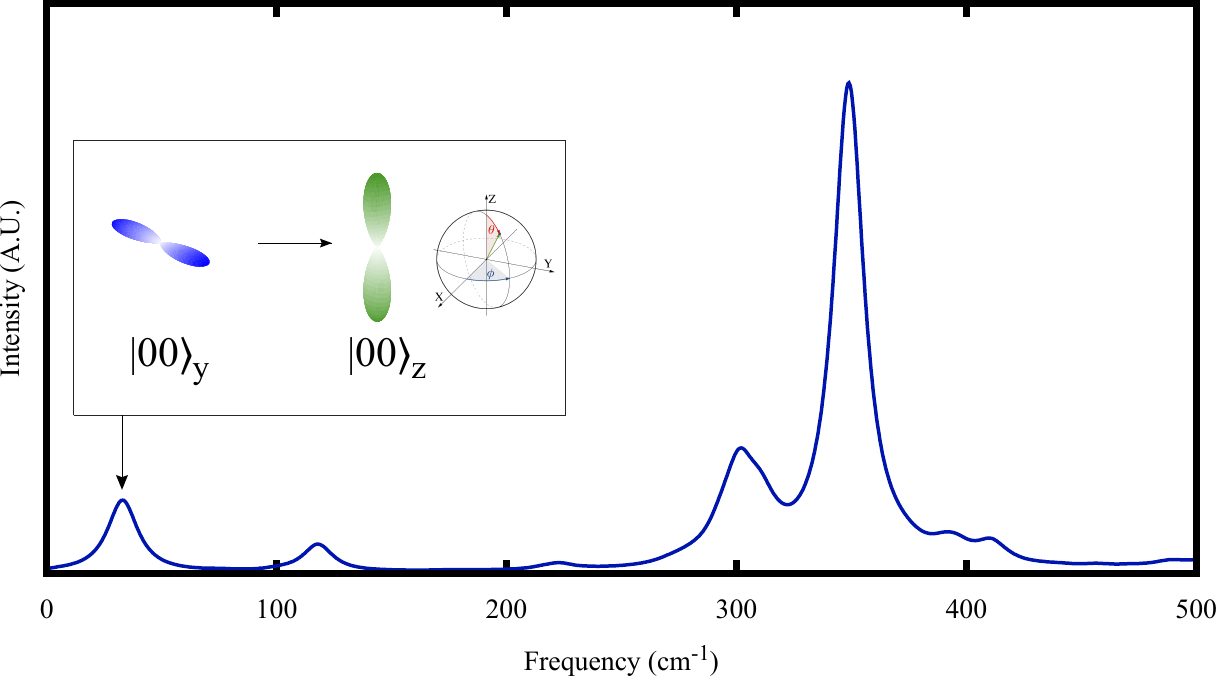}
\end{center}
\caption{Raman Spectrum at P$=800$ cm$^{-1}$ for D$_2$ in case 4 at 300 K. The reorientaional mode between the ground state and a meta-stable state pointing along the z axis (($\ket{00}_y \: \rightarrow  \ket{00}_z$) is indicated by the arrow. The mode appears at low Raman shift in this case as the two orientations are close in energy.}
\label{fig:ReorientRaman2}
\end{figure}

As in the previous cases we see a breaking of the $2J+1$ rotor degeneracy as P increases. This also means that as the $m_J$ degeneracy is lifted, transitions which in the zero potential limit are associated with the elastic Rayleigh scattering ($\Delta J=0$), acquire a small non-zero frequency and are strongly Raman active. These transitions gradually emerge from the Rayleigh line as the strength of the potential is increased from $P=0\rightarrow100$ (see orange triangles in Fig. \ref{fig:Raman4}). They represent a more subtle re-orientation associated with the axis of rotation of a weakly hindered rotor with fixed angular momentum. At high field strengths a distinct type of re-orientational modes appear where the molecule transitions from one oscillator minimum to another as shown Fig. \ref{fig:ReorientRaman2}. In Fig. \ref{fig:Raman4}, indicated with green triangles, one of these modes can be seen emerging as a shoulder from the $S_0(0)$ ($J=0 \rightarrow 2$) peak and gradually moving to lower frequency with increasing P. These modes are Raman active at intermediate field strengths but the predicted Raman intensity gradually decreases as P is increased.

\section{Conclusion}
\label{sec:Conclusion}

We have presented a detailed analysis of Raman spectra for a diatomic molecule in four different simple potentials composed of spherical harmonics. For weak potentials the eigenstates describe a nearly-free rotor, and the good quantum numbers are $J$ and $m_J$. With strong potential they describe molecules in well defined orientations, with harmonic oscillations (librons) where two SHO quantum numbers are associated with each inequivalent direction. The eigenstates in each direction are described by precisely two quantum numbers since each minimum has two degrees of freedom. For intermediate potential strength, the eigenstates are mixed, with no good quantum numbers. In addition, for homoatomic molecules, all states are symmetric to nuclear exchange, which creates a parity degeneracy at high field.

We can identify three types of Raman active modes: excitations of rotational states, excitations of oscillator states, and re-orientational excitations. Oscillator and rotor excitations have well defined selection rules, but all transitions between mixed states are Raman active. Consequently, as the potential increases, the simple roton spectra at low potential changes to a complex manifold containing many overlapping states. The roton, libron and reorientation modes all have different mass dependencies. Experimentally this can serve as a tool to identify the character of each mode by taking ratios of Raman peak frequencies between different isotopes.

Nuclear exchange symmetry is not explicit in the model, but appears implicitly through the well defined parity of the wavefunctions. It is interesting to note that in the harmonic oscillator limit the parity does not appear as a Raman selection rule because each SHO level is a doublet with odd and even parity. This suggests that at large field strength, changes in the Raman spectra no longer reflect the spin symmetry of the hydrogen molecule, therefore these spectra are agnostic to the breakdown of ortho- and para- states observed in NMR studies at intermediate pressures \cite{Meier2020}.

We note that even our most complex potential, case 4, has higher symmetry than known diatomic crystalline materials. For example, molecular hydrogen adopts an hcp phase which would be modeled by a more complicated mean-field potential that has one minimum in the c-direction, and six minima with molecules tilted slightly out of the basal plane. The Raman spectra shown here are simpler than could ever be obtained experimentally, however this allows us to easily understand and interpret the character of the different transitions involved \footnote{A combination of experimental geometry and strong preferred orientation can make some modes unobservable}.  

\section{Data availability}
Codes can be obtained from the authors on request. No experimental data was collected in this study.
\section{Acknowledgements}
GJA acknowledges the support of the European Research
Council Grant Hecate Reference No. 695527 and a Royal Society Wolfson fellowship. EPSRC funded studentships for PICC and IBM and computing time (UKCP grant P022561).
\bibliographystyle{elsarticle-num} 
\bibliography{main}

\end{document}